# Entanglement of Pure Two-Mode Gaussian States

R. W. Rendell and A. K. Rajagopal
Naval Research Laboratory, Washington DC  20375

The entanglement of general pure Gaussian two-mode states is examined in terms of the coefficients of the quadrature components of the wavefunction. The entanglement criterion and the entanglement of formation are directly evaluated as a function of these coefficients, without the need for deriving local unitary transformations.  These reproduce the results of other methods for the special case of symmetric pure states which employ a relation between squeezed states and Einstein-Podolsky-Rosen correlations. The modification of the quadrature coefficients and the corresponding entanglement due to application of various optical elements is also derived.

## I. Introduction

Continuous quantum variables have emerged as an alternative to discrete-level systems for performing quantum information processing tasks.  Gaussian states play a central role because they can be readily produced from reliable sources [1] and controlled experimentally using accessible sets of operations such as beam splitters, phase shifters and squeezers [2] and efficient detection systems. They can be formed in a variety of physical systems including light field modes [3], cold atoms [4], and excitons in photonic cavities [5].  Entanglement between two Gaussian modes is now routinely generated in the laboratory, e.g. two output beams of a parametric down converter are sent through optical fibers [1] or in atomic ensembles interacting with light [6]. Such continuous variable entanglement can be efficiently produced using squeezed light and linear optics [3].  Two-mode entangled states can enhance the capability of two parties to communicate as well as other applications.  Gaussian states have already been utilized in realizations of quantum key distributions [7], teleportation [8] and electromagnetically induced transparency [9].

Progress has also been made theoretically in the understanding of quantum entanglement in continuous spaces between two or more modes.  Separability criteria have been established for general two-mode continuous variable states [10, 11].  For the special case of two-mode Gaussians, this becomes a necessary and sufficient criterion for entanglement [10, 11] and this result has been extended to arbitrary bipartite Gaussian states [12].  To go beyond entanglement criteria and quantify the amount of entanglement in the state requires an entanglement measure. For general bipartite pure states (continuous or discrete), the von Neumann entropy of either of the reduced states is, under reasonable assumptions, a unique entanglement measure [13]: $E_{vN}(\psi) = S(tr_i(|\psi\rangle\langle\psi|))$.  Physically, it corresponds to the fraction of maximally entangled states that can be obtained by local operations and classical communication (LOCC) when applied to an ensemble of such states.  The von Neumann entropy for the simplest example of continuous variable states, the two-mode squeezed states, is well-known [14] and has been used in a variety of applications:

$$E_{vn}(\psi_{Sq}) = \cosh^2 r \ln(\cosh^2 r) - \sinh^2 r \ln(\sinh^2 r) \qquad (1)$$

where r is the two-mode squeezing parameter and is not dependent on the squeezing phase.

Both the squeezed states and the Einstein-Podolsky-Rosen (EPR) variables have played key roles in obtaining both entanglement criteria and entanglement measures for more general states. Separability criteria have been established in terms of inequalities of the EPR variables [10,11] and these have been used to experimentally demonstrate continuous variable entanglement [3,4]. It has been found [15-17] that any pure bipartite Gaussian can be locally transformed into products of two-mode squeezed states so that its entanglement can be evaluated via eq.(1). Entanglement measures for mixed states are more complicated and involve carrying out a nontrivial optimization over all possible decompositions of the density matrix into pure states. Although there are several proposed entanglement measures for mixed states [18], a physically appealing generalization of the von Neumann entropy is the entanglement of formation [19]:

$$E_F(\rho) = \inf\left\{\sum_k p_k E_{vN}(\psi_k)\right\}, \text{ for all (possibly continuous) decompositions } \rho = \sum_k p_k |\psi_k\rangle\langle\psi_k|.$$

This reduces to $E_{vN}$ for the case of pure states but is in general difficult to calculate because of the optimization involved in finding the infimum. For continuous variable mixed states, $E_F$ has been derived only for two-mode Gaussians which are symmetric in the interchange of the two modes [20, 21]. In this case the optimization was made possible by locally transforming the covariance matrix to standard form [10,11] and using a special property of the two-mode squeezed states, namely that they have the smallest entanglement of all symmetric states with a given value of the EPR correlation. As with the case of bipartite pure Gaussians, entanglement of symmetric mixed two-mode Gaussians is obtained via eq.(1) after performing specific local transformations. This property of the squeezed states and EPR correlations has also been used to find lower bounds on $E_F$ for more general Gaussians [22]. Some results have also been obtained for another entanglement measure of two mode Gaussians, the logarithmic negativity [23,24].

In this work, we examine the entanglement properties of general pure two-mode Gaussians directly in terms of the coefficients of the quadrature modes. For the important case of pure two-mode Gaussians, our results recover the known answer without the need to find the local unitary transformations required in the method of [15-17] for bipartite pure Gaussians or, in the case of two-mode symmetric Gaussians, the method of [20,21]. Section II characterizes the general two-mode Gaussian in terms of the quadrature coefficients and presents various expectation values required for the entanglement criterion and the reduced density matrices necessary for calculating the von Neumann entropy. Section III analyzes the entanglement criteria in terms of these coefficients. Section IV derives the entanglement of formation in terms of the quadrature components by finding the eigenvalues of an effective Hamiltonian associated with the two-mode Gaussians. For the special case of symmetric states, these results are related to the mixed-state expression for the entanglement of formation and that of the associated squeezed states. Section V discusses how the two-mode Gaussian coefficients are modified by Gaussian operations and the subsequent changes in entanglement. Section VI presents concluding remarks.

## II. Characterization of Pure Two-Mode Gaussians

Two spatially separated quantum modes i=1,2 can be fully described by means of field quadratures [25], the amplitude quadrature, $\hat{X}_i = (\hat{a}_i^\dagger + \hat{a}_i)/\sqrt{2}$, and the phase quadrature, $\hat{Y}_i = i(\hat{a}_i^\dagger - \hat{a}_i)/\sqrt{2}$, in analogy to the position, $\hat{q}_i$, and momentum, $\hat{p}_i$ of the original EPR variables. Here $\hat{a}_i, \hat{a}_i^+$ are the destruction and creation operators of mode i, obeying the usual commutation rules (we use units with $\hbar = 1$). The amplitude and phase quadratures, which determine the properties of the optical beams both as to entanglement and polarization correlations, are routinely measured [3,4]. In order to study pure two-mode entanglement in terms of the coefficients of the wave function, we examine the general normalized two-mode Gaussian wave function in the representation of the amplitude quadrature or, equivalently, the position variables:

$$\Psi(q_1, q_2) = N \exp-(\alpha q_1^2 + \beta q_2^2 + 2\gamma q_1 q_2)/2 \qquad (2)$$

where $\alpha = \alpha_1 + i\alpha_2$, $\beta = \beta_1 + i\beta_2$, $\gamma = \gamma_1 + i\gamma_2$, $N^2 = \Delta/\pi$, and $\Delta^2 = \alpha_1 \beta_1 - \gamma_1^2 > 0$. Terms linear in $q_i$ were not included in eq.(2) because these correspond to displacements which do not affect the entanglement.

The bipartite Gaussian states and density matrices correspond to a special class in that the covariance matrix determines them completely. They are characterized by the first and second moments of the canonical operators, despite the underlying infinite dimensional Hilbert space. For two modes, the 4x4 covariance matrix contains all the necessary information to determine its entanglement properties for both entanglement criteria and entanglement measures. The 4x4 covariance V matrix [10-12] is written in terms of the 2x2 partitioned matrices A, B, C, and $C^T$:

$$V = \begin{pmatrix} A & C \\ C^T & B \end{pmatrix}; \quad A = \begin{pmatrix} \langle \hat{q}_1^2 \rangle & \langle \{\hat{q}_1 \hat{p}_1\} \rangle \\ \langle \{\hat{q}_1 \hat{p}_1\} \rangle & \langle \hat{p}_1^2 \rangle \end{pmatrix}, \quad B = \begin{pmatrix} \langle \hat{q}_2^2 \rangle & \langle \{\hat{q}_2 \hat{p}_2\} \rangle \\ \langle \{\hat{q}_2 \hat{p}_2\} \rangle & \langle \hat{p}_2^2 \rangle \end{pmatrix},$$

$$C = \begin{pmatrix} \langle \hat{q}_1 \hat{q}_2 \rangle & \langle \hat{q}_1 \hat{p}_2 \rangle \\ \langle \hat{p}_1 \hat{q}_2 \rangle & \langle \hat{p}_1 \hat{p}_2 \rangle \end{pmatrix}; \quad C^T = \begin{pmatrix} \langle \hat{q}_1 \hat{q}_2 \rangle & \langle \hat{q}_2 \hat{p}_1 \rangle \\ \langle \hat{p}_2 \hat{q}_1 \rangle & \langle \hat{p}_1 \hat{p}_2 \rangle \end{pmatrix}.$$

(3)

where $\hat{p}_i = -i\partial/\partial q_i$ is the operator conjugate to $q_i$, i =1, 2 and $\langle \{\hat{a}\hat{b}\} \rangle = \langle \hat{a}\hat{b} + \hat{b}\hat{a} \rangle / 2$.

These are explicitly calculated in terms of the coefficients of the wavefunction in eq.(2) as:

$$\langle q_1^2 \rangle = \beta_1/(2\Delta^2); \quad \langle q_2^2 \rangle = \alpha_1/(2\Delta^2); \quad \langle q_1 q_2 \rangle = -\gamma_1/(2\Delta^2); \qquad (4a)$$

$$\langle p_1^2 \rangle = \frac{\beta_1 |\alpha|^2 - \alpha_1(\gamma_1^2 - \gamma_2^2) - 2\gamma_1\gamma_2\alpha_2}{2\Delta^2};$$

$$\langle p_2^2 \rangle = \frac{\alpha_1 |\beta|^2 - \beta_1(\gamma_1^2 - \gamma_2^2) - 2\gamma_1\gamma_2\beta_2}{2\Delta^2}; \qquad (4b)$$

$$\langle p_1 p_2 \rangle = \frac{(\alpha_1\gamma_1 + \alpha_2\gamma_2)\Delta^2 + (\alpha_1\gamma_2 - \alpha_2\gamma_1)(\alpha_1\beta_2 - \gamma_1\gamma_2)}{2\alpha_1\Delta^2};$$

$$\langle \{q_1 p_1\} \rangle = \frac{(\gamma_1\gamma_2 - \alpha_2\beta_1)}{2\Delta^2}; \quad \langle q_1 p_2 \rangle = \frac{(\gamma_1\beta_2 - \gamma_2\beta_1)}{2\Delta^2};$$

$$\langle q_2 p_1 \rangle = \frac{(\gamma_1\alpha_2 - \gamma_2\alpha_1)}{2\Delta^2}; \quad \langle \{q_2 p_2\} \rangle = \frac{(\gamma_1\gamma_2 - \alpha_1\beta_2)}{2\Delta^2}. \qquad (4c)$$

These will be used in Section II in the discussion of the entanglement criterion and in Section III in the discussion of the entanglement of formation. Calculation of the entanglement of formation also requires the two 1 - mode marginal density matrices of the 2 - mode wave function:

$$\langle q_1|\hat{\rho}_I|q_1'\rangle = \int_{-\infty}^{\infty} dq_2 \Psi(q_1,q_2)\Psi^*(q_1',q_2) =$$
$$= \frac{\Delta}{\sqrt{\pi\beta_1}} \exp-\frac{1}{4\beta_1}\left\{\begin{array}{l}[2\alpha_1\beta_1 - \gamma_1^2 + \gamma_2^2](q_1^2 + q_1'^2) \\ + 2i[\alpha_2\beta_1 - \gamma_1\gamma_2](q_1^2 - q_1'^2) \\ -2[\gamma_1^2 + \gamma_2^2]q_1 q_1'\end{array}\right\} \quad (5)$$

and similarly:

$$\langle q_2|\hat{\rho}_{II}|q_2'\rangle = \frac{\Delta}{\sqrt{\pi\alpha_1}} \exp-\frac{1}{4\alpha_1}\left\{\begin{array}{l}[2\beta_1\alpha_1 - \gamma_1^2 + \gamma_2^2](q_2^2 + q_2'^2) \\ + 2i[\beta_2\alpha_1 - \gamma_1\gamma_2](q_2^2 - q_2'^2) \\ -2[\gamma_1^2 + \gamma_2^2]q_2 q_2'\end{array}\right\} \quad (6)$$

## III. Entanglement Criterion

We evaluate the separability criterion for continuous variable two-mode states derived by Simon [10] using the wavefunction of eq.(2). This criterion involves the determinants of the covariance matrix and the partitioned submatrices in eq.(3) and these can be evaluated using eq.(4):

$$\det A = \langle q_1^2\rangle\langle p_1^2\rangle - \langle\{q_1 p_1\}\rangle^2 = \frac{\alpha_1\beta_1 + \gamma_2^2}{4\Delta^2} = \frac{1}{4} + \frac{\gamma_1^2 + \gamma_2^2}{4\Delta^2}$$
$$= \det B = \langle q_2^2\rangle\langle p_2^2\rangle - \langle\{q_2 p_2\}\rangle^2; \quad \det C = -\frac{\gamma_1^2 + \gamma_2^2}{4\Delta^2}; \quad (7)$$
$$\det V = \det A \det B + (\det C)^2 - Tr = 1/16$$

where

$$Tr \equiv tr(AJCJBJC^T J) = \frac{(\gamma_1^2 + \gamma_2^2)}{8\Delta^2}\left[1 + \frac{(\gamma_1^2 + \gamma_2^2)}{\Delta^2}\right] \quad (8)$$

and

$$J = \begin{pmatrix} 0 & 1 \\ -1 & 0 \end{pmatrix}.$$

The expressions for det A and det B represent the variances for each of the modes which enter the one-mode Heisenberg uncertainty relations. These would equal ¼ for minimum uncertainty states. That these deviate from the minimum uncertainty is due to the entanglement of the two one-mode states which is present for nonzero values of the coefficient γ in eq.(2). However, the amount of entanglement depends not only on γ but also on α and β as well, as will be shown in Section III. Note that the two 1-mode density matrices in eqs.(5) and (6) are distinct as expected from the one-mode dispersions of position given in eq.(4a). This reflects the fact that eq.(2) represents a generally asymmetric pure two-mode Gaussian. In spite of this, the uncertainty relations given in eq.(7) show that the two systems have the same uncertainty content.

The 2-mode Heisenberg uncertainty relation [10] is expressed in the form

$$\Omega_2^2 = \det A \det B + \left(\frac{1}{4} - \det C\right)^2 - \frac{1}{4}(\det A + \det B) - Tr \qquad (9)$$
$$\equiv \det V + \frac{1}{16} - \frac{1}{4}(\det A + \det B + 2\det C) \geq 0.$$

From eqs.(7,8), it is seen that $(\det A + \det B + 2\det C) = 1/2$ so that the equality sign is obtained for the state in eq.(2), as is expected for a pure Gaussian state.

The Simon separability condition [10] is valid for arbitrary two-mode continuous variable states and is stated in terms of the quantity:

$$E_S = \det A \det B + \left(\frac{1}{4} - |\det C|\right)^2 - Tr - \frac{1}{4}(\det A + \det B)$$
$$= \det V + \frac{1}{16} - \frac{1}{4}(\det A + \det B + 2|\det C|) \qquad (10)$$
$$\geq 0$$

This becomes a necessary and sufficient condition for bipartite Gaussian states and thus also serves as an entanglement criterion [10]. For the two-mode Gaussian of eq.(2), this is found from eqs.(7,8) to be:

$$E_S = -\frac{(\gamma_1^2 + \gamma_2^2)}{4\Delta^2} = \det C \ . \qquad (11)$$

This shows that the two-mode wave function is always entangled for nonzero values of $\gamma$. That the sign of detC determines the entanglement of the two-mode Gaussian was already shown by Simon [10].

## IV. Entanglement of Formation

For the pure state of eq.(2) considered here, the entanglement of formation can be obtained by calculating the eigenvalues and evaluating the von Neumann entropy. In order to discuss the results in a physical way, it is important to recall the corresponding results for a one-mode Gaussian system. The following construction is inspired by different ways of obtaining the Gaussian density matrix in the literature [26-28]. We take the approach that a density matrix may be generated by a Hermitian Hamiltonian. While this is not the most general way to realize a given density matrix, it leads to a practical procedure. We therefore consider the following biquadratic Hamiltonian generated by the canonically conjugate Hermitian operators $\hat{Q}, \hat{P}$ obeying the standard commutation rules, $[\hat{Q}, \hat{P}] = i$ and others zero:

$$\hat{H} = \frac{1}{2}\left(D\hat{Q}^2 + E\hat{P}^2 + F(\hat{Q}\hat{P} + \hat{P}\hat{Q})\right), \tag{12}$$

where D, E, F are real parameters due to Hermiticity of the Hamiltonian. Introduce the 1-mode squeezing operators such that:

$$\hat{A} = \lambda\hat{Q} + i\mu\hat{P},\ \hat{A}^\dagger = \lambda^*\hat{Q} - i\mu^*\hat{P} \tag{13}$$

where $\lambda, \mu$ are two complex numbers such that $[\hat{A},\hat{A}^\dagger] = 1 = \lambda\mu^* + \lambda^*\mu$, and the other commutators zero. The standard parameterization in terms of a squeezing parameter, r, and phase $\varphi$ is:

$$\lambda = (Coshr + e^{i\varphi} Sinhr)/\sqrt{2},\ \mu = (Coshr - e^{i\varphi} Sinhr)/\sqrt{2}. \tag{14}$$

We find that eq.(12) can be recast into the standard form

$$\hat{H} = \omega\left(\hat{A}^\dagger\hat{A} + \frac{1}{2}\right) \tag{15}$$

provided we identify the parameters in the following way:

$$\omega|\lambda|^2 = D/2;\ \omega|\mu|^2 = E/2;\ i\omega(\lambda^*\mu - \lambda\mu^*) = F. \tag{16}$$

Clearly we have the identity

$$\omega^2 = DE - F^2. \tag{17}$$

We now find the eigenstates of the Hamiltonian in terms of the usual number representation by working in the representation where $\hat{Q}|x\rangle = x|x\rangle,\ \hat{P} = -i\,\partial/\partial x$. These are:

$$\langle x|n\rangle = \psi_n\left(x/|\mu|\sqrt{2}\right)\exp{-i\left(F/4\omega|\mu|^2\right)x^2}$$
$$\hat{H}|n\rangle = \omega\left(n + \frac{1}{2}\right),\ n = 0,1,2,\cdots. \tag{18}$$

The eigenfunctions in eq.(18) are the orthonormal harmonic oscillator wave functions. We now use Mehler's formula

$$\sum_{m=0}^{\infty} z^m \psi_n(x)\psi_n^*(y) = \frac{1}{\sqrt{1-z^2}}\exp{-\left[\left(\frac{1+z^2}{1-z^2}\right)\left(\frac{x^2+y^2}{2}\right) - 2\left(\frac{z}{1-z^2}\right)xy\right]} \tag{19}$$

to express the density matrix in the form:

$$\langle x|\hat{\rho}_1|x'\rangle = \frac{\langle x|\exp-\kappa\omega(\hat{A}^\dagger\hat{A}+1/2)|x'\rangle}{Z} =$$

$$= \frac{1}{|\mu|\sqrt{2\pi}}\left(\frac{1-e^{-\kappa\omega}}{1+e^{-\kappa\omega}}\right)^{1/2}\exp(-iF(x^2-x'^2)/4|\mu|^2\omega) \qquad (20)$$

$$\exp-\frac{1}{2|\mu|^2}\left\{\left(\frac{1+e^{-2\kappa\omega}}{1-e^{-2\kappa\omega}}\right)\left(\frac{x^2+x'^2}{2}\right)-\left(\frac{2e^{-\kappa\omega}}{1-e^{-2\kappa\omega}}\right)xx'\right\}$$

Equivalently we express this in terms of the eigenvalues of the density matrix:

$$\langle x|\hat{\rho}_1|x'\rangle = (1-e^{-\kappa\omega})\sum_{m=0}^{\infty}(e^{-\kappa\omega})^m \psi_m(x)\psi_m^*(x'). \qquad (20')$$

Expressing this in the form of the marginal 1-mode density matrices given in eqs.(5,6), we have:

$$\langle x|\hat{\rho}_1|x'\rangle = \frac{\sqrt{a-c}}{\sqrt{\pi}}\exp-\frac{1}{2}\{a(x^2+x'^2)-2cxx'\}\exp-\frac{ia'}{2}(x^2-x'^2) \qquad (21)$$

with:

$$a = \frac{1}{2|\mu|^2}\left(\frac{1+e^{-2\kappa\omega}}{1-e^{-2\kappa\omega}}\right); \; c = \frac{1}{2|\mu|^2}\left(\frac{2e^{-\kappa\omega}}{1-e^{-2\kappa\omega}}\right); \; a' = \frac{F}{2\omega|\mu|^2} \qquad (22)$$

Calculating the three variances associated with this density matrix results in:

$$\langle \hat{x}^2\rangle = \frac{1}{2(a-c)}; \; \langle \hat{p}^2\rangle = \frac{(a+c)}{2}+\frac{a'^2}{2(a-c)};$$

$$\langle \{\hat{x}\,\hat{p}\}\rangle = -\frac{a'}{2(a-c)}. \qquad (23)$$

The Heisenberg relationship is then:

$$\Omega_{H1}^2 = \frac{1}{4}\frac{(a+c)}{(a-c)} = \frac{1}{4}\left(\frac{1+e^{-\kappa\omega}}{1-e^{-\kappa\omega}}\right)^2 \qquad (24)$$

The von Neumann entropy associated with this density matrix is

$$S = -Tr\hat{\rho}_1\ln\hat{\rho}_1 \qquad (25)$$

and using eqs.(20') and (24) we find

$$S = (\Omega_{H1}+1/2)\ln(\Omega_{H1}+1/2)-(\Omega_{H1}-1/2)\ln(\Omega_{H1}-1/2). \qquad (26)$$

Equation (24) shows an interesting relationship between the parameters in the Hamiltonian and the density matrix and the expectation values of the quadratures:

$$\kappa \omega = \ln\left(\frac{\Omega_{H1} + 1/2}{\Omega_{H1} - 1/2}\right). \tag{27}$$

Comparing the marginal one-mode density matrices obtained from the two-mode system with the above, we obtain $a_{1A} = (2\alpha_1\beta_1 - \gamma_1^2 + \gamma_2^2)/2\beta_1$; $c_{1A} = (\gamma_1^2 + \gamma_2^2)/2\beta_1$. Using eq.(24), we have:

$$\Omega_{HA}^2 = \frac{1}{4} + \frac{\gamma_1^2 + \gamma_2^2}{4\Delta^2} = \frac{1}{4}\left(\frac{\alpha_1\beta_1 + \gamma_2^2}{\alpha_1\beta_1 - \gamma_1^2}\right) \tag{28}$$

The results for $\Omega_{HB}^2$ turn out to be the same as $\Omega_{HA}^2$, as is expected for a pure state, even though the marginal density matrices in eqs.(5,6) are different. We therefore find that the two marginal von Neumann entropies are the same and determine the entanglement of formation for the two–mode Gaussian wave function in terms of the quadrature coefficients as:

$$E_F = (\Omega_{HA} + 1/2)\ln(\Omega_{HA} + 1/2) - (\Omega_{HA} - 1/2)\ln(\Omega_{HA} - 1/2) \tag{29}$$

Note that $E_F = 0$ for vanishing coupling between the modes, $\gamma = 0$ in eq.(2), as would be expected. However, the magnitude of the entanglement also generally depends on the values of α and β, but only through their real parts.

In the special case of a two-mode squeezed state, the quadrature coefficients in eq.(2) take the form $\alpha = \beta = (1 + \lambda^2)/(1 - \lambda^2)$ and $\gamma = -2\lambda/(1 - \lambda^2)$, where the squeezing parameter r and phase φ are defined by $\lambda = -\exp(i\varphi)\tanh r$. For this particular symmetric Gaussian, the quadrature coefficients obey the relation $\alpha^2 - \gamma^2 = 1$. From eq(28), the two-mode squeezed state has $\Omega_{HA} = \cosh 2r/2$ so that the entanglement of formation in eq.(29) reduces to the known squeezed state expression in eq.(1). Note that the squeezed state entanglement depends only on the squeezing parameter r and is independent of the phase, φ.

Eqs.(28) and (29) can also be compared with the symmetric mixed state expression for $E_F$ from Giedke et al [20, 21] in the case of symmetric pure two-mode Gaussians. The mixed-state result relies on applying local unitary transformations (which do not modify the entanglement) to the covariance matrix in eq.(3) to bring it into standard form [29]:

$$V_S = \frac{1}{2}\begin{pmatrix} n & 0 & k_q & 0 \\ 0 & n & 0 & k_p \\ k_x & 0 & m & 0 \\ 0 & k_p & 0 & m \end{pmatrix} \tag{30}$$

For the case of n=m, which is the case of symmetric two-mode Gaussians, Giedke et al are able to perform the infimum in the definition of the entanglement of formation for the state corresponding to eq.(30). The result is that $E_F$ for the general symmetric (i.e. n=m) two-mode

Gaussian in Eq.(30) is equal to the entanglement of the special two-mode squeezed state having the same value of EPR correlation as the state in eq.(30). This implies that the entanglement can be obtained from eq.(1) with a special squeezing parameter equal to $r = \ln[(n-k_q)(n+k_p)]/2$.

We can compare this to the present result if we restrict eqs.(28, 29) to the special case of symmetric states (i.e. α=β) and restrict the Giedke et al result to pure states. A symmetric pure state requires n=m and $\det V_S = 1/16$ in eq.(30), so that $k_p = -k_x$ and $n^2 - k_x^2 = 1$. Using these results in the expression for r leads to $n = \cosh 2r$. Using this in eq.(1) and comparing with eq.(29), we find that the entanglement of Giedke et al can also be described by eq.(29) with $\Omega_{HA} = n/2$. The value of n in eq.(30) is then related to the quadrature coefficients of eq.(2) by $n = 1 + |\gamma|^2 / \Delta^2$. The entanglement of symmetric pure two-mode Gaussians can be found either by eq.(1) with special value of r after transforming V into standard form eq.(30) or directly using eq.(29).

## V. Gaussian Operations on Entanglement

The ability to accurately manipulate Gaussian quantum states by means of optical elements leads to possibilities of quantum information processing using continuous variables [7-9]. Understanding the effect of operations on Gaussian states and the resulting affect on the corresponding entanglement is therefore of interest. Of particular concern are Gaussians operations (with classical communication) [30], which are just the experimentally accessible set of operations that can be realized with optical elements such as beam splitters, phase shifts and squeezers, together with homodyne measurements. The effect of optical elements on the entanglement of two-mode Gaussians has been studied in terms of logarithmic negativity [23]. In the following, we examine how such transformations affect the quadrature coefficients of the pure-state wave function of eq.(2) and thereby the entanglement of formation from eqs. (28, 29).

The pure two-mode Gaussian wave function in eq.(2) may be considered as having been generated in a number of ways by applying various operations to the vacuum [31]. For example, the state resulting from application of a two–mode squeezing, $\hat{S}_{12} = \exp-\left(\zeta_{12}\hat{a}_1^\dagger \hat{a}_2^\dagger - \zeta_{12}^* \hat{a}_1 \hat{a}_2\right)$, $\zeta_{12} = r_{12} e^{i\varphi_{12}}$, followed by a one-mode squeezing, $\hat{S}_1 = \exp-\frac{1}{2}\left(\zeta_1 \hat{a}_1^{\dagger 2} - \zeta_1^* \hat{a}_1^2\right)$, $\zeta_1 = r_1 e^{i\varphi_1}$ [25] to the vacuum is equivalent to eq.(2) to within local unitary transformations [15-17]. Here the two complex parameters, $\zeta_{12}$ and $\zeta_1$, represent the squeezings. It is sufficient to apply the one-mode squeezing to only one of the two modes. The squeezed mode annihilation operators, $\tilde{a}_i = \hat{S}_1 \hat{S}_{12} \hat{a}_i \hat{S}_{12}^\dagger S_1^\dagger$, acting on the squeezed vacuum lead to the equations:

$$[(q_1(1+\lambda_1) + ip_1(1-\lambda_1))\cosh r_1 + (q_2 - ip_2)\lambda_{12}]\psi(q_1, q_2) = 0$$
$$[(q_1(1+\lambda_1^*) - ip_1(1-\lambda_1^*))\lambda_{12} \cosh r_1 + (q_2 + ip_2)]\psi(q_1, q_2) = 0$$
(31)

where the squeezing parameters are defined as $\lambda_j = e^{i\varphi_j} \tanh r_j$, j=1 and j=12. The wave function is the quadrature amplitude representation of the squeezed vacuum, $\psi(q_1, q_2) = \langle q_1, q_2 | \hat{S}_1 \hat{S}_{12} | 0,0 \rangle$, since $0 = a_1 a_2 |0,0\rangle = \tilde{a}_1 \tilde{a}_2 (\hat{S}_1 \hat{S}_{12} |0,0\rangle)$. From the solution to these differential equations, it is then found that α, β, and γ in eq.(2) are related to the squeezing parameters as follows:

$$\alpha = \frac{(1+\lambda_1) + \lambda_{12}^2(1+\lambda_1^*)}{(1-\lambda_1) - \lambda_{12}^2(1-\lambda_1^*)}, \quad \beta = \frac{(1-\lambda_1) + \lambda_{12}^2(1-\lambda_1^*)}{(1-\lambda_1) - \lambda_{12}^2(1-\lambda_1^*)},$$

$$\gamma = \frac{2\lambda_{12}\sqrt{(1-|\lambda_1|^2)}}{(1-\lambda_1) - \lambda_{12}^2(1-\lambda_1^*)}. \tag{32}$$

From eqs.(28, 29) and (32), we find that the entanglement of formation of eq.(2) in this example can be expressed in terms of both the one and two–mode squeezing parameters. The most general case of eq.(2) can be obtained if local unitaries were applied in addition to the squeezing operators. Eq.(32) would then be further modified by parameters describing the unitaries in addition to the one and two-mode squeezing parameters. In the special case when the one–mode squeezing is turned off, $r_1 = 0$, we obtain a symmetric Gaussian wave function with α=β. This corresponds to the two-mode squeezed state with entanglement described by eq.(1). This is equivalent to the general symmetric pure Gaussian by local unitaries [15-17]. If the two-mode squeezing is turned off instead, $r_{12}=0$, we obtain an asymmetric but unentangled Gaussian given by the product of two single-mode squeezed states.

The state in this example could be further modified by using additional operators representing various optical elements. For example, a beam splitter and a phase plate can be described by [32]:

$$\hat{B}_{12} = \begin{pmatrix} e^{-i\phi}\cos\theta/2 & -e^{-i\phi}\sin\theta/2 \\ \sin\theta/2 & \cos\theta/2 \end{pmatrix} \tag{33}$$

Consider applying this to the two-mode squeezed state, then the requirement that the operator $\hat{B}_{12}\hat{S}_{12}\hat{a}_i\hat{S}_{12}^\dagger\hat{B}_{12}^\dagger$ annihilate the vacuum again leads to differential equations analogous to eq.(31) which can be solved to find the new quadrature coefficients representing two-mode squeezing, beam splitting, and a phase shift:

$$\alpha = \frac{1 + 2\tilde{\lambda}\cos\theta + \tilde{\lambda}^2}{1-\tilde{\lambda}^2}, \quad \beta = \frac{1 - 2\tilde{\lambda}\cos\theta + \tilde{\lambda}^2}{1-\tilde{\lambda}^2}, \quad \gamma = \frac{2\tilde{\lambda}\cos\theta}{1-\tilde{\lambda}^2} \tag{34}$$

where $\tilde{\lambda} = \lambda_{12}e^{i\phi}$. The entanglement resulting from these operations can be calculated using eqs.(28,29). If single-mode squeezings were also applied to each mode, then the effective annihilation operator become $\hat{B}_{12}S_2\hat{S}_1\hat{S}_{12}\hat{a}_i\hat{S}_{12}^\dagger S_1^\dagger S_2^\dagger \hat{B}_{12}^\dagger$, and the quadrature components and entanglement could again be calculated in a similar way. In this case, eq.(34) would be modified by six additional parameters due to the single-mode squeezings. Ref. [23] had examined the entangling capability of passive optical elements in terms of the logarithmic negativity and focused on two-mode Gaussians. The present results allow a similar study in terms of the entanglement of formation for arbitrary two-mode Gaussians.

# VI. Concluding Remarks

This work examined the entanglement properties for general pure two-mode Gaussian states directly in terms of the quadrature coefficients of the wave function. The entanglement criterion and the entanglement of formation were evaluated as a function of these coefficients. This direct calculation eliminates the need for first deriving local unitary transformations to put the covariance matrix into standard form. The results were compared to those for symmetric Gaussians which relate two-mode squeezed states and Einstein-Podolsky-Rosen correlations. The modification of the quadrature coefficients and the corresponding entanglement due to application of various optical elements was also derived. A discussion of the effects of dissipation or noise on the two-mode system is given in [33].

Acknowledgements: We thank the Office of Naval Research for partial support of this work.